\documentclass[modern]{aastex63}

\usepackage{amsmath}
\usepackage{amssymb}
\usepackage{rotating}

\usepackage{xcolor}

\newcommand\fss{f_{\rm s}}

\newcommand\tE{t_{\rm E}}
\newcommand\teff{t_{\rm eff}}

\newcommand\piEN{\pi_{\textrm{E},N}}
\newcommand\piEE{\pi_{\textrm{E},E}}

\newcommand\tpar{t_{\rm 0,par}}
\newcommand\muvec{\boldsymbol{\mu}}

\newcommand\muS{\boldsymbol{\mu}_{\rm source}}
\newcommand\muL{\boldsymbol{\mu}_{\rm lens}}

\newcommand\piE{\pi_{\rm E}}
\newcommand\piEvec{\boldsymbol{\pi_{\rm E}}}

\newcommand\model{\boldsymbol{\omega}}
\newcommand\data{\boldsymbol{d}}
\newcommand\mf{\boldsymbol{\alpha}}
\newcommand\tabdots{\multicolumn{1}{c}{\dots}}

\received{}
\revised{}
\accepted{}

\shorttitle{Mass function of isolated stellar remnants}
\shortauthors{P. Mr\'oz et al.}

\begin{document}

\title{Measuring the mass function of isolated stellar remnants with gravitational microlensing. II. Analysis of the OGLE-III data}

\correspondingauthor{Przemek Mr\'oz}
\email{pmroz@astro.caltech.edu}

\author[0000-0001-7016-1692]{Przemek Mr\'oz}
\affil{Division of Physics, Mathematics, and Astronomy, California Institute of Technology, Pasadena, CA 91125, USA}
\affil{Astronomical Observatory, University of Warsaw, Al. Ujazdowskie 4, 00-478 Warszawa, Poland}

\author[0000-0001-5207-5619]{Andrzej Udalski}
\affil{Astronomical Observatory, University of Warsaw, Al. Ujazdowskie 4, 00-478 Warszawa, Poland}

\author[0000-0002-9658-6151]{\L{}ukasz Wyrzykowski}
\affil{Astronomical Observatory, University of Warsaw, Al. Ujazdowskie 4, 00-478 Warszawa, Poland}

\author[0000-0002-2335-1730]{Jan Skowron}
\affil{Astronomical Observatory, University of Warsaw, Al. Ujazdowskie 4, 00-478 Warszawa, Poland}

\author[0000-0002-9245-6368]{Rados\l{}aw Poleski}
\affil{Astronomical Observatory, University of Warsaw, Al. Ujazdowskie 4, 00-478 Warszawa, Poland}

\author[0000-0002-0548-8995]{Micha\l{} K. Szyma\'nski}
\affil{Astronomical Observatory, University of Warsaw, Al. Ujazdowskie 4, 00-478 Warszawa, Poland}

\author[0000-0002-7777-0842]{Igor Soszy\'nski}
\affil{Astronomical Observatory, University of Warsaw, Al. Ujazdowskie 4, 00-478 Warszawa, Poland}

\author[0000-0001-6364-408X]{Krzysztof Ulaczyk}
\affil{Department of Physics, University of Warwick, Coventry CV4 7 AL, UK}
\affil{Astronomical Observatory, University of Warsaw, Al. Ujazdowskie 4, 00-478 Warszawa, Poland}

\begin{abstract}
Our knowledge of the birth mass function of neutron stars and black holes is based on observations of binary systems but the binary evolution likely affects the final mass of the compact object. Gravitational microlensing allows us to detect and measure masses of isolated stellar remnants, which are nearly impossible to obtain with other techniques. Here, we analyze a sample of 4360 gravitational microlensing events detected during the third phase of the OGLE survey. We select a subsample of 87 long-timescale low-blending events. We estimate the masses of lensing objects by combining photometric data from OGLE and proper-motion information from OGLE and \textit{Gaia}~EDR3. We find 35 high-probability dark lenses -- white dwarfs, neutron stars, and black holes -- which we use to constrain the mass function of isolated stellar remnants. In the range $1-100\,M_{\odot}$, occupied by neutron stars and black holes, the remnant mass function is continuous and can be approximated as a power-law with a slope of $0.83^{+0.16}_{-0.18}$ with a tentative evidence against a broad gap between neutron stars and black holes. This slope is slightly flatter than the slope of the mass function of black holes detected by gravitational wave detectors LIGO and Virgo, although both values are consistent with each other within the quoted error bars. The measured slope of the remnant mass function agrees with predictions of some population synthesis models of black hole formation.
\end{abstract}

\keywords{Gravitational microlensing (672), Stellar remnants (1627), Black holes (162), Neutron stars (1108), White dwarf stars (1799)}

\section{Introduction} \label{sec:intro}

Detecting and directly measuring masses of isolated stellar remnants, especially neutron stars and black holes, is virtually impossible with traditional astrophysical methods. Our knowledge of the mass function of neutron stars and black holes is based on observations of binary systems but the binary evolution likely affects the final mass of the compact object. However, isolated neutron stars and black holes must be ubiquitous in our Galaxy. Knowledge of their mass function would give us important clues about the evolution of massive stars, core collapse and supernova mechanisms, etc.

Masses of neutron stars in binary systems can be measured with precise timing observations of radio pulsars either in double neutron star or neutron star-white dwarf systems. Mass measurements are also possible for neutron stars in X-ray binaries by combining X-ray and optical observations \citep[e.g.,][]{ozel2016}. Masses of neutron stars in double neutron-star systems peak at $1.33\pm 0.09\,M_{\odot}$, whereas those in neutron star-white dwarf binaries are more massive (typically $1.54\pm 0.23\,M_{\odot}$) \citep[e.g.,][]{kiziltan2013}. The maximum observed mass of a neutron star is about $2.14\,M_{\odot}$ \citep{cromartie2020}.

All known stellar-mass black holes were found in binary systems -- either in black hole-star (via radial velocity or in X-ray binaries) or black hole--black hole/neutron star binaries found via gravitational waves by LIGO and Virgo. The distribution of dynamical masses of black holes in X-ray binaries is consistent with a narrow Gaussian at $7.8 \pm 1.2\,M_{\odot}$ \citep{ozel2010} with an apparent absence of compact objects in the $2-5\,M_{\odot}$ range (the so-called ``mass gap''; \citealt{ozel2010,farr2011}). The distribution of masses of black holes in 47 compact-binary mergers from the second LIGO--Virgo Gravitational-Wave Transient Catalog \citep{gwtc2020} is consistent with a broken power law or a power law with a Gaussian feature. According to that study, the minimum black hole mass is lower than $6.6\,M_{\odot}$ (with 90\% credibility). \citet{belczynski2012} and \citet{fryer2012} proposed that the mass gap may be caused by the supernova explosion mechanism that should be driven by instabilities with a rapid growth time. This hinders formation of compact objects with intermediate masses. However, some ``mass-gap'' objects may still be formed, for example, from mergers of neutron stars and white dwarfs.

Indeed, recent discoveries indicate that objects with masses intermediate between those of neutron stars and black holes do exist. The product of the binary neutron star merger in GW170817 has a mass of $2.74^{+0.04}_{-0.01}\,M_{\odot}$ \citep{abbott2017}. LIGO and Virgo have also detected a coalescence of a massive black hole with a $2.50-2.67\,M_{\odot}$ ``mass-gap'' object in gravitational-wave signal GW190814 \citep{abbott2020}. \citet{thompson2019} and \citet{jayasinghe2021} discovered $\sim 3\,M_{\odot}$ dark companions orbiting giant stars.

Isolated dark stellar remnants may be detected in gravitational microlensing events \citep[e.g.,][]{paczynski1996,gould2000,mao2002,bennett2002}. However, lens mass measurements are possible only in special cases, when the Einstein timescale $\tE$, the microlens parallax $\piE$, and the relative lens-source proper motion $\mu$, are known:
\begin{equation}
M = \frac{\tE\mu}{\kappa\piE},
\label{eq:mass}
\end{equation}
where $\kappa=8.144\,\mathrm{mas\,yr}^{-1}$. The values of $\tE$ and $\piE$ can be measured (or constrained) from the light curve of the event, whereas $\mu$ is usually unknown. However, the most probable distribution of $\mu$ can be inferred from the Milky Way models, which allows us to estimate the masses and distances to lensing objects. This method was first proposed by \citet{wyrzykowski2016} and \citet{wyrzykowski2020b}, who searched for stellar remnants in OGLE microlensing data. However, as we explain in \citet{mroz2021}, the masses of compact objects inferred by \citet{wyrzykowski2020b} are overestimated and their ``mass-gap'' and black hole events are, in fact, most likely due to main-sequence stars, white dwarfs, or neutron stars.

In this paper, we re-analyze a large sample of microlensing events detected in the third phase of the OGLE survey with the main aim of searching for stellar remnant candidates and measuring their mass function.

\section{Data}

\subsection{Event selection}

The photometric data analyzed in this paper were collected during the years 2001--2009 during the third phase of the Optical Gravitational Lensing Experiment (OGLE-III) survey \citep{udalski2003}. We selected 91 fields with the largest number of epochs, covering an area of about 31\,deg$^2$ of the Galactic bulge. The vast majority of the collected images (up to $\sim 2500$ per field) were taken through the $I$-band filter, closely resembling that of the standard Cousins filter. A smaller number of exposures (1 to 35 per field) were collected in the $V$-band. OGLE-III used a mosaic CCD camera with a field of view of 0.34\,deg$^2$ mounted on the 1.3-m Warsaw Telescope located at Las Campanas Observatory, Chile. Thanks to the small pixel scale ($0.26"$ per pixel) and superb sky conditions (typical seeing $1-1.5"$), OGLE-III could detect objects as faint as $I\approx 21$ in 120-s exposures in dense regions of the Galactic bulge.

Several studies used OGLE-III observations of the Galactic bulge to search for gravitational microlensing events. Over 4000 events were discovered in real-time by the OGLE Early Warning System \citep[EWS;][]{udalski2003}. The system was designed for detection of ongoing microlensing events. \citet{wyrzykowski2015} selected a sample of 3718 standard events found in the OGLE-III data (of which 1409 had not been detected before by EWS), which they used to construct maps of the mean Einstein ring crossing time and compared them with predictions of Milky Way models. Additional 59 long-timescale events exhibiting an annual microlens parallax effect were selected by \citet{wyrzykowski2016}, who searched for stellar remnant (white dwarf, neutron star, and black hole) candidates. 

In this paper, we analyze 3620 ``class A'' microlensing events detected by \citet{wyrzykowski2015, wyrzykowski2016}. In addition, we run the event finder algorithm of \citet{mroz2017} on OGLE-III data and find an extra 740 events. Thus, our final sample comprises 4360 events. 

To select stellar remnant candidates, we apply several selection cuts. First, we expect that microlensing events due to black holes have relatively long timescales because $\tE \propto \sqrt{M}$. Long-timescale events are likely to exhibit light curve deviations caused by the orbital motion of Earth (the so-called annual microlens parallax effect). Even if the amplitude of the effect is too small to be reliably measured from the light curve, its value may be tightly constrained by the light curve data, which also provides useful information. Second, we use the proper motion of the source to infer the lens properties and so we select events for which the majority of the light comes from the source star (so that the proper motion of the source can be approximated as the proper motion of the baseline object, which we measure from the archival OGLE data). This is quantified by the dimensionless blending parameter $\fss$, which is the ratio of the source flux to the total unlensed flux of the event.

In the first step, we fit all 4360 light curves with a standard point-source point-lens microlensing model with parallax. There may be up to four possible solutions describing every light curve due to inherent degeneracies \citep[e.g.,][]{smith2003,gould2004,skowron2011}. Then, we select events with at least one solution with $\tE \geq 60$\,d and $\fss \geq 0.8$ and remove binary-lens or binary-source events, as well as events with incomplete, poorly-sampled, or low-amplitude light curves. We end up with 87 events with timescales between 60 and 300\,d. In this timescale range, the detection efficiency is virtually constant. We extract the optimized light curves of selected events with the difference image analysis method \citep{alard1998,wozniak2000}.

\subsection{Proper motions}

Out of 87 long-timescale low-blending events in our sample, proper motions of only 46 are available in the \textit{Gaia} Early Data Release 3 \citep[EDR3;][]{gaia2016,gaia_edr3}. However, it is known that the completeness of \textit{Gaia}~EDR3 is reduced in crowded areas such as the Galactic bulge \citep{gaia_edr3,fabricius2021}. Some sources in crowded regions may have spurious astrometric solutions and their proper-motion measurements may suffer from catastrophic errors \citep[e.g.,][]{hirao2020,mroz2021}.

\begin{figure}
\centering
\includegraphics[width=0.8\textwidth]{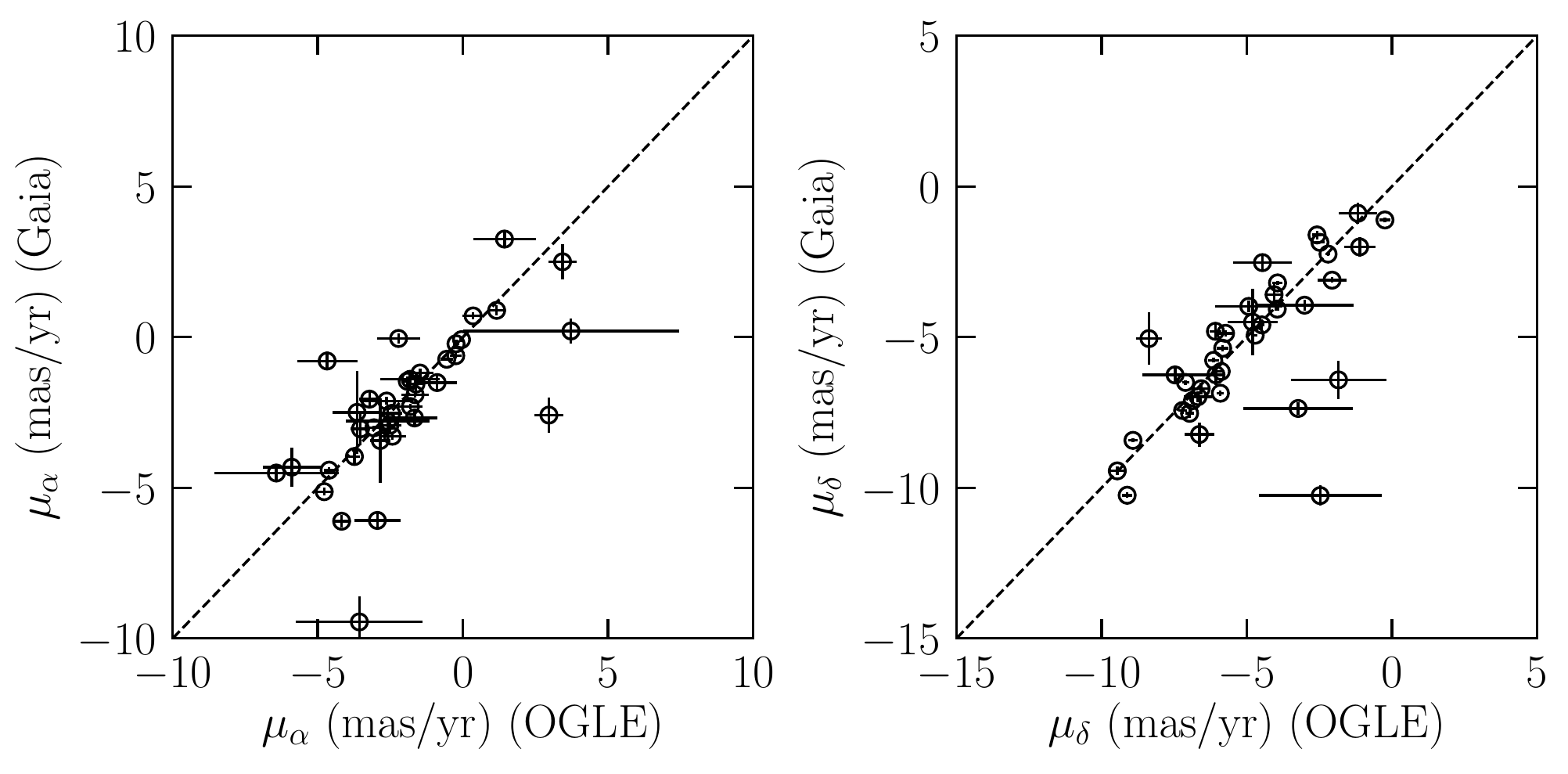}
\caption{Comparison between OGLE and \textit{Gaia} proper motions for 39 common events.}
\label{fig:pm}
\end{figure}

Precise measurements of proper motions are also possible with long-term ground-based observations. We use proper-motion measurements calculated using observations collected by the fourth phase of the OGLE survey \citep[OGLE-IV; 2010--2020;][]{udalski2015}. Positions of stars are measured on individual frames using the astrometric OGLE pipeline and are tied to the \textit{Gaia}~EDR3 reference frame. A detailed description of the OGLE Uranus astrometry project will be published elsewhere (Udalski et al. 2021, in preparation). OGLE proper motions are available for 68 events, 39 of which are common with \textit{Gaia}~EDR3. Figure~\ref{fig:pm} presents the comparison between OGLE and \textit{Gaia} proper motions of these common stars, which agree well. They are listed in Table~\ref{tab:pm}.

In the following analysis, we use OGLE proper motions for 68 events. If OGLE measurements are not available and \textit{Gaia}~EDR3 astrometric solution has the renormalized unit weighted error \citep{lindegren2021} smaller than 1.4, we use \textit{Gaia} (6 events). For the remaining 13 events, we assume that their proper motion is consistent with that of Galactic bulge stars $(\mu_l,\mu_b)=(-6.12,-0.19) \pm 2.64$~mas\,yr$^{-1}$. This proper motion corresponds to the velocity of the Sun relative to the Milky Way center \citep{schonrich2010} as seen from the distance of 8\,kpc, and the uncertainty corresponds to the typical velocity dispersion in the Galactic bulge (100\,km\,s$^{-1}$).

\section{Methods}

A detailed description of how to estimate the lens mass given the event light curve and the source proper motion is presented by \citet{mroz2021}. We estimate the masses of lenses using Equation~\ref{eq:mass}. The values of $\tE$ and $\piE$ are measured (or constrained) from the light curve model, whereas $\mu$ is unknown -- its value may be only constrained based on prior information from the Milky Way model. Note that $\mu = |\muvec| = |\muL-\muS|$, where $\muL$ and $\muS$ are proper motions of the lens (which is unknown) and the source (which may be measured by OGLE or \textit{Gaia}), respectively. Moreover, if the microlens parallax is detected in the light curve of the event, the direction of $\muvec \propto \piEvec/\piE$ is also known.

Our event models have eight parameters. Five of them are ``standard'' point-lens point-source microlensing parameters that describe the shape of the light curve. These are: time $t_0$ and separation $u_0$ (in Einstein radius units) during the closest lens-source approach, effective timescale of the event $\teff = \tE|u_0|$, and North and East components of the microlens parallax vector $\piEN$ and $\piEE$. Two parameters ($\mu_{s,N}$ and $\mu_{s,E}$) describe the North and East components of the source proper motion vector (relative to the solar system barycenter) and are measured by either OGLE or \textit{Gaia}. The final parameter is the relative lens-source proper motion $\mu$, its value is constrained only by the Milky Way model. Here, we assume that the source is located at a distance of 8\,kpc in the Galactic bulge, we use the Milky Way model from \citet{mroz2021} and use the Kroupa mass function as our priors. As discussed by \citet{mroz2021}, the choice of these priors has little effect on the inferred lens mass and distance. In particular, we opt not to use \textit{Gaia} parallaxes to estimate source distances as they are not accurate enough to provide meaningful constraints. Model parameters (except $\mu_{s,N}$ and $\mu_{s,E}$) are measured in a geocentric frame that is moving with a velocity equal to that of the Earth at a fiducial time $\tpar$. 

Every light curve may have up to four degenerate solutions (differing by signs of $u_0$, $\piEN$, and $\piEE$). Moreover, the lens may be located either in the Galactic disk or in the bulge, so the distribution of $\mu$ may be bimodal (as shown in Figure~1 of \citealt{mroz2021}). To handle possible multiple solutions, we derive posterior probability distributions with the nested sampling Monte Carlo algorithm MLFriends \citep{buchner2019} using UltraNest\footnote{\url{https://johannesbuchner.github.io/UltraNest/}} \citep{buchner2021}. In the nested sampling algorithms, the entire eight-dimensional parameter space is filled with a set of live points taken from prior distributions (in the present case, we use uniform priors for all parameters but $\mu_{s,N}$ and $\mu_{s,E}$, which are taken from Gaussian distributions). Then the live point with the lowest likelihood is removed from the set and replaced with a new one on the condition that its likelihood is larger than the likelihood of the removed point, so that the volume sampled by the live points shrinks at every iteration. The removed points are weighted by their likelihood and stored and then are used to generate the posterior distribution for all parameters. We run the sampler with a minimum of 1000 live points throughout the run and terminate the integration when the sum of weights of live points is smaller than 0.05 (frac\_remain) of the sum of weights of accepted points.

The main advantage of our approach is that we can simultaneously explore all possible solutions. Standard Markov chain Monte Carlo samplers (for example, \textsc{emcee} by \citet{foreman2013}) may not work well if the posterior is multi-modal. To test our algorithm, we derived posterior distributions of masses and distances to all lenses from our sample with \textsc{emcee} and the results were very similar for 83 of 87 events. For the remaining four events, we re-run UltraNest with a larger number of 2000 live points and set frac\_remain to 0.01 and obtained virtually identical posterior distributions of parameters as in our initial models. In all four cases, \textsc{emcee} did not properly sample the multi-modal posterior.

For every analyzed event we derive a posterior distribution in the mass -- distance space (Table~\ref{tab:all}). We use the empirical mass--absolute brightness relations for main-sequence stars\footnote{\url{http://www.pas.rochester.edu/~emamajek/EEM_dwarf_UBVIJHK_colors_Teff.txt}} \citep{pecaut2013} and interstellar extinction maps of \citet{nataf2013} to derive the expected distribution of $I$-band brightness of the lens. The extinction varies with the distance -- we assume that the extinction is proportional to the integrated density of interstellar material along the line of sight following the model of \citet{sharma2011} and we normalize it to \citet{nataf2013} extinction maps. We compare the expected $I$-band brightness with the blended flux from the microlensing model. If a putative main-sequence lens is brighter than the blend, this indicates that the lens is dark. For every event, we calculate the probability $p$ that the lens is not luminous.

\begin{figure}
\centering
\includegraphics[width=.7\textwidth]{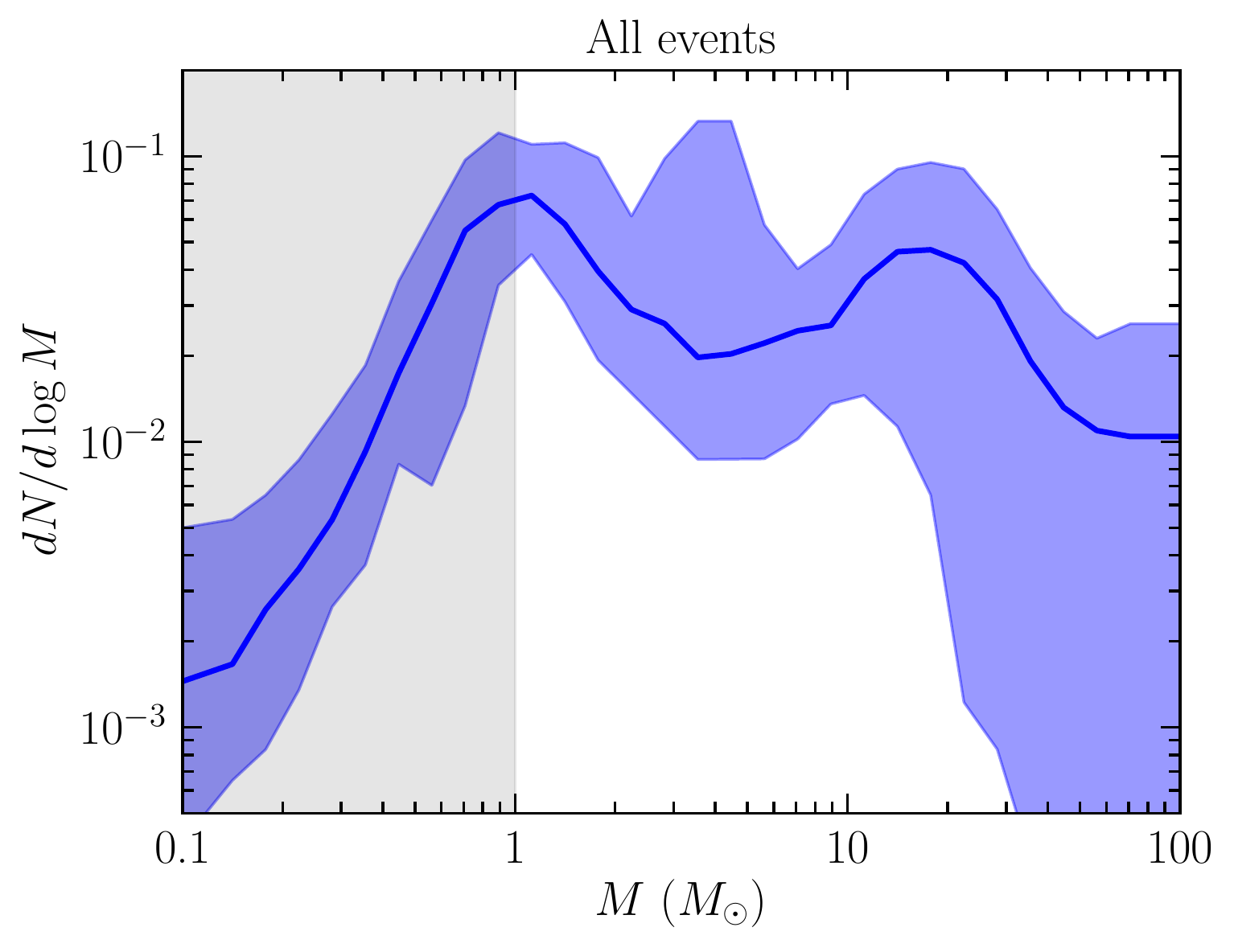} \\
\includegraphics[width=.7\textwidth]{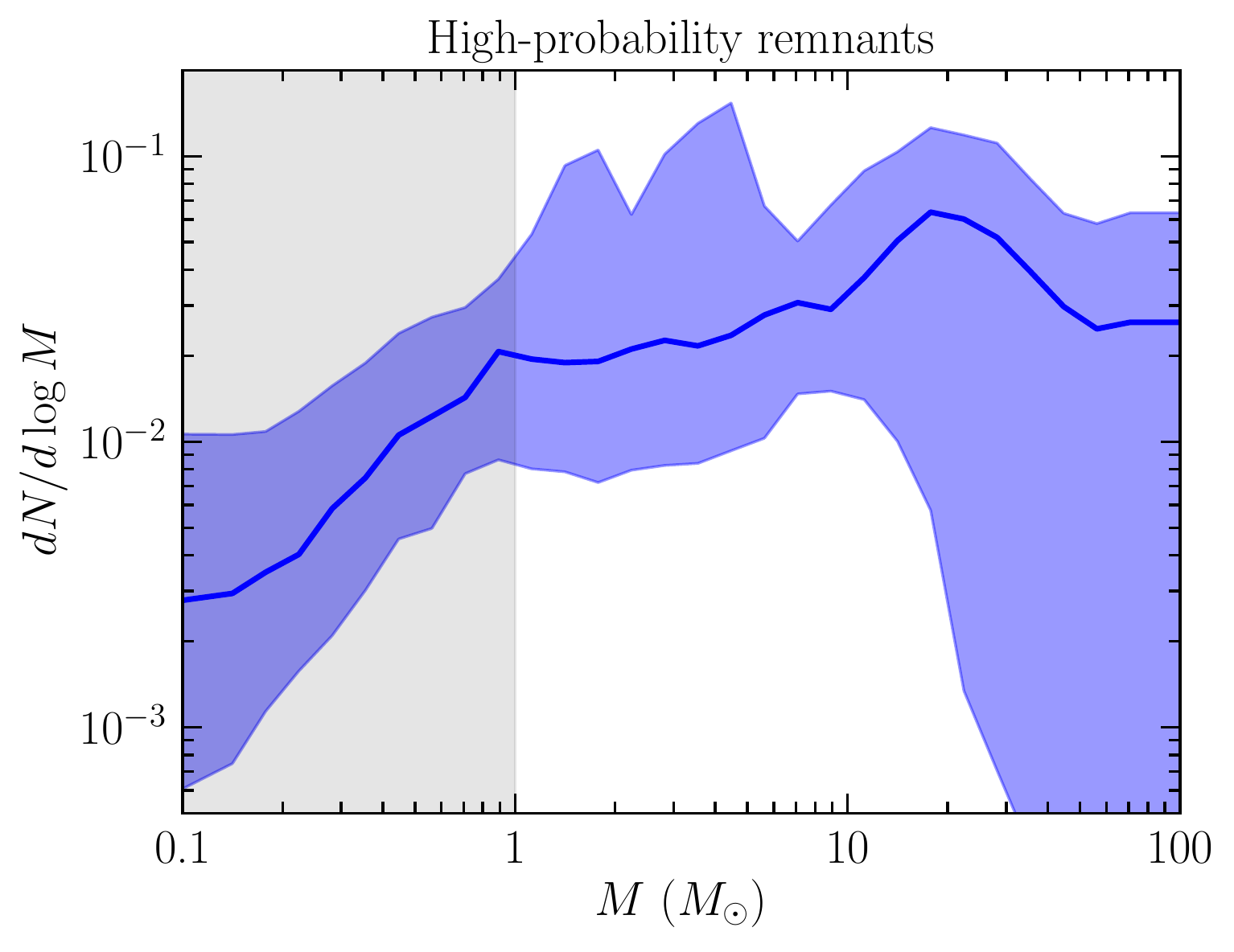}
\caption{Upper panel: Distribution of lens masses in the analyzed sample of microlensing events. The mass function is approximated as a histogram with 30 bins in $\log M$ with a width of 0.1\,dex each. The blue shaded area represents a 68\% credibility region and the solid blue line marks a median of the posterior distribution for bin heights. Lower panel: Distribution of masses of high-probability stellar remnants. The analyzed sample of lenses is incomplete for $M<1\,M_{\odot}$ so the shape of the histogram does not reflect the real shape of the remnant mass function below $1\,M_{\odot}$.}
\label{fig:mf}
\end{figure}

\section{Lens mass function}

Masses of individual lenses in our sample may be determined with large uncertainties, the posterior mass distributions may be asymmetric or even bimodal in some cases. We use a hierarchical Bayesian modeling \citep{hogg2010} to infer the mass function of lenses.

We have a sample of $N$ events. For the $n$th event, we derive the posterior distribution for eight parameters $p(\model_n | \data_n)$ using nested sampling, where $\data_n$ are data for that event. This distribution is calculated using the prior distribution $p_0(\model_n)$, which includes information from the Milky Way model and a fiducial mass function of lenses $g_0(M)$ (the calculation of the prior is described in detail by \citealt{mroz2021}, we use the Kroupa mass function as a prior on the mass function $g_0(M)$).
For every event, we use nested sampling to obtain a set of $K_n$ samples $\model_{nk}$ which represents a random draw from the posterior distribution.

Let us now assume that the mass function of lenses $f_{\mf} (M) = dN/dM$ can be described by a set of parameters $\boldsymbol{\alpha}$. The likelihood function $\mathcal{L}_{\mf}$ for parameters $\mf$ is
\begin{equation}
\mathcal{L}_{\mf} = p(\{\data_n\}_{n=1}^N|\mf) = \prod_{n=1}^N \int d\model_n p(\data_n | \model_n) p (\model_n | \mf),
\label{eq:l_alpha}
\end{equation}
where $p(\model_n | \mf) = f_{\mf} (M) p_0(\model_n) / g_0(M)$ \citep{hogg2010}. This step is crucial for the inference of the mass function -- we replace the fiducial mass function $g_0(M)$ with the function we aim to model $f_{\mf} (M)$. Thus, the derived mass function does not depend on $g_0(M)$. The integral in Equation (\ref{eq:l_alpha}) may be approximated as the sum over samples from the posterior:
\begin{equation}
\mathcal{L}_{\mf} \propto \prod_{n=1}^N \frac{1}{K_n} \sum_{k=1}^{K_n} \frac{f_{\mf} (M_{nk})}{g_0(M_{nk})}.
\end{equation}
We implicitly assume here that all lenses are drawn from the same mass function. This may not be true in general, for example, the remnant mass functions may be different in the Galactic disk and bulge. The analyzed here sample of events is too small to reliably separate these two populations.

In our primary model, the mass function can be approximated as a histogram with $B$ bins in $\log M$:
\begin{equation}
f_{\mf}(M) = \frac{dN}{dM}= \frac{1}{M \ln 10}\frac{dN}{d\log M} = \frac{1}{M \ln 10} \sum_{b=1}^B \exp(\alpha_b) s_b(\log M),
\end{equation}
where $s_b$ is the step function and $\sum_{b=1}^B \exp(\alpha_b) = 1$. Following \citet{hogg2010}, we assume a smoothness prior on $\mf$: $p(\mf) = \exp\left(-\sum_{b=2}^B (\alpha_b-\alpha_{b-1})^2\right)$ \citep{kitagawa1996}.

We also consider a simpler model, in which the lens mass function can be expressed as a broken power law:
\begin{equation}
f_{\mf}(M) = \frac{dN}{dM}= \begin{cases}
a_0 M^{-\alpha_0} & \text{for $0.1\,M_{\odot} < M < M_{\rm break}$},\\
a_1 M^{-\alpha_1} & \text{for $M_{\rm break} < M < 100\,M_{\odot}$},
\end{cases}
\end{equation}
where $a_0$ and $a_1$ are normalization constants, we assume flat priors on $\alpha_0$ and $\alpha_1$.

In both cases, we use the Markov chain Monte Carlo sampler \textsc{emcee} \citep{foreman2013} to derive the posterior distributions for the mass function parameters $\mf$.

\section{Results and discussion}
\label{sec:discussion}

We derive posterior probability distributions for all 87 long-timescale low-blending events in our sample and then we fit the hierarchical model to derive the mass function of lenses. We approximate the mass function as a histogram with 30 bins in $\log M$ with a width of 0.1\,dex each. The constraints on the mass function are presented in the upper panel of Figure~\ref{fig:mf}, the shaded region represents the 68\% credibility interval and the solid blue line marks the median of the posterior distribution of $\mf$. 

The mass function peaks around $1\,M_{\odot}$; this peak is a selection effect, however. Note that the analyzed sample of microlensing events contains only events with timescales longer than $\tE=60$\,d and so the mass function is biased toward larger masses (since $\tE\propto\sqrt{M}$). For a comparison, we also measure the combined mass function for events with timescales longer than $\tE=80$\,d. Both mass functions match well for masses greater than $1\,M_{\odot}$ but the latter contains fewer low-mass lenses. Thus, our combined distribution reflects the real mass function of lenses for $M\gtrsim 1\,M_{\odot}$ but the peak and the turnover for lower masses are just a selection effect.

When we fit a broken power-law model, we find the mass function slopes $\alpha_0=-0.80^{+0.56}_{-0.73}$ for $0.1 < M < 1\,M_{\odot}$ and $\alpha_1=1.34^{+0.15}_{-0.12} $ for $1 < M < 100\,M_{\odot}$. It is also clear from Figure~\ref{fig:mf} that for masses larger than $\approx 20\,M_{\odot}$, the data do not have enough constraining power and the allowed credible region is large (we can provide only upper limits on the mass function in that mass range).

We then select high-probability stellar remnants. In our sample, there are 35 events with a probability that the lens is dark $p>0.95$, 27 events with $p>0.98$, and 21 events with $p>0.99$. Our constraints on the mass function of high-probability dark lenses are presented in the lower panel of Figure~\ref{fig:mf}. Among 35 events with $p>0.95$, 25 objects have their proper motions measured either from OGLE or \textit{Gaia}, proper motions of 10 objects are not constrained. We checked, however, that the combined mass function of 25 events with known proper motions is very similar to that presented in Figure~\ref{fig:mf}. The sample of high-probability stellar remnants contains mostly faint events, which explains the lack of OGLE/\textit{Gaia} proper motions.

The shape of the mass function of high-probability dark lenses does not resemble that of the mass function of the entire sample. We measure the power-law slopes of $\alpha_0=-0.51^{+0.95}_{-1.64}$ for $0.1 < M < 1\,M_{\odot}$ and $\alpha_1=0.83^{+0.16}_{-0.18} $ for $1 < M < 100\,M_{\odot}$. When we restrict the sample to events with proper motions measured by either OGLE or \textit{Gaia}, we find $\alpha_0=-1.50^{+1.52}_{-2.07}$ and $\alpha_1=0.92^{+0.22}_{-0.20} $, respectively. The mass function slope in the range $1 < M < 100\,M_{\odot}$ is slightly flatter than the slope of the black hole mass function ($1.58^{+0.82}_{-0.86}$) inferred from the second LIGO-Virgo Gravitational-Wave Transient Catalog \citep{gwtc2020} (their broken power-law model), although both values are consistent with each other within the quoted error bars.

We can also compare the measured slope with the theoretical predictions based on population synthesis calculations by \citet{olejak2020} using the StarTrack code \citep{belczynski2002,belczynski2008}. They consider isolated black holes that are formed as a result of single star evolution, disruptions of binary star systems, or mergers of compact objects. \citet{olejak2020} provide synthetic catalogs of black holes separately in the Galactic disk and bulge. We fit a power law model to their simulated data in the range $6-20\,M_{\odot}$ and find the slope of $0.9$ and $2.2$ for the Galactic disk and bulge populations, respectively. The former is consistent with our findings but the latter slope is steeper. Our sample contains objects located both in the Galactic disk and bulge but its size is too small to reliably separate these two populations.

In the range $3-10\,M_{\odot}$, occupied by ``mass-gap'' objects and black holes, our remnant mass function is continuous with no evidence for a gap between neutron stars and black holes. This result should be treated with caution. Masses of individual lenses have relatively large uncertainties (typically $0.3-0.5$\,dex in $\log M$) so one can argue that we cannot detect a narrow feature (gap) in the remnant mass function. We thus run simulations in which we assume a log-uniform mass function in the ranges $1-2\,M_{\odot}$ and $6-30\,M_{\odot}$. We draw a sample of 35 objects from that fiducial mass function and assign each mass measurement the uncertainty of 0.05, 0.3, and 0.5\,dex. We then use the hierarchical Bayesian modeling to infer the mass function based on the simulated data. 

\begin{figure}
\centering
\includegraphics[width=.8\textwidth]{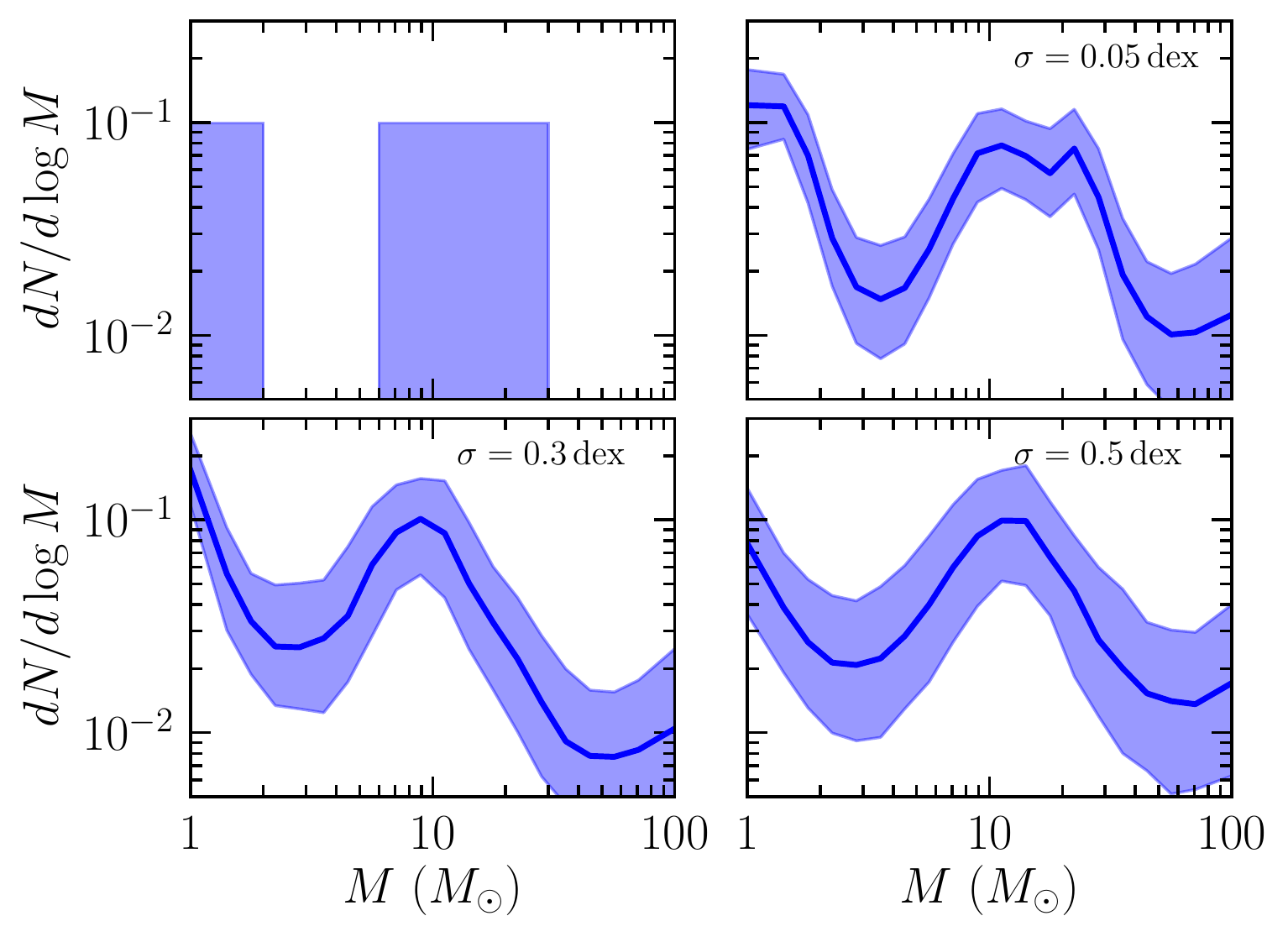} \\
\caption{We run simulations to check if we can recover a mass gap ($2-6\,M_{\odot}$) in the log-uniform mass function (upper left panel) of remnants. We draw 35 events from our fiducial mass function and assign each mass measurement the uncertainty of 0.05, 0.3, and 0.5\,dex. We use our hierarchical modeling to infer the mass function using the simulated data. We are able to recover the gap in all cases.}
\label{fig:sim}
\end{figure}

Results of our simulations are presented in Figure~\ref{fig:sim}. We are able to recover the gap in all cases, although the shape of the mass function becomes more blurry and the credible intervals become larger as the uncertainties increase. Nonetheless, we find that the shape of the mass function is much better constrained if the sample of simulated events is larger. We thus plan to analyze a larger sample of microlensing events detected during the OGLE-IV phase, which contains four times more events than the current sample. This will enable us to provide stronger constraints on the mass function in the ``mass gap'' regime.

We cannot exclude that the mass gap is only partly filled with objects that form from mergers of binary neutron stars \citep[e.g.,][]{abbott2017}. Another source of contamination may be close binary systems of compact objects -- if the orbital separation is much smaller than the size of the Einstein ring (typically a few au), such a system can be regarded as effectively a single lens for microlensing. The contamination from close binary main-sequence stars is less likely. We re-computed the dark companion probability assuming an equal-mass binary lens instead of a single star lens, which is higher than 88\% for all objects classified as high-probability remnants (higher than 95\% for 31/35 events).

The main limitation of our work is the assumption that remnants and normal stars share the same velocity distribution. However, neutron stars may receive large natal kicks at birth \citep{hobbs2005}, while there is no agreement about natal kicks of black holes (e.g., \citealt{callister2020} and references therein). If the proper motion of the lens is high enough, the Einstein timescale may be shorter than our threshold of 60\,days and the event is not included in our sample. Moreover, large natal kicks may affect the determination of the lens mass, as discussed in more detail by \citet{mroz2021}. The amplitude of effect depends on the geometry of individual events, location of the lens, as well as the poorly known distribution of kick velocities.

In the future, thanks to advances in precise astrometry and interferometry, it may be possible to directly measure masses (as well as velocities) of individual isolated stellar remnants. 
Direct mass measurements for many events will become possible thanks to precise astrometric observations by the \textit{Gaia} satellite \citep{rybicki2018} and its planned successors \citep{hobbs2021}. A new path for measuring masses of isolated objects is opened up by the first resolution of microlensed images by the GRAVITY interferometer \citep{dong2019}. Although now interferometric observations are possible only for the brightest events, the planned upgrades to the GRAVITY instrument will enable observations of dozens of fainter events\footnote{\url{https://www.mpe.mpg.de/7480772/GRAVITYplus_WhitePaper.pdf}}. Further in the future, the planned \textit{Nancy Grace Roman Telescope} is expected to detect hundreds of microlensing events by isolated black holes \citep{penny2019}. \textit{Roman} will provide both precise photometry and astrometry, enabling us to directly measure the mass function of isolated stellar remnants. 

\section*{Acknowledgements}

We thank Jim Fuller, Ilya Mandel, and Aleksandra Olejak for discussions and their comments on the manuscript.
This work benefited from help of Krzysztof Rybicki, Katarzyna Kruszy\'nska, as well as students at the Astronomical Observatory, University of Warsaw, particularly: Zofia Kaczmarek, Kornel Howil, Mateusz Bronikowski, and Anna Jab\l{}onowska.
P.M.~acknowledges the support from Heising-Simons Foundation Grant \#2018-1036 awarded to J.~Fuller. \L{}.W.~acknowledges the support from the Polish National Science Center grants: Harmonia No.~2018/30/M/ST9/00311 and Daina No.~2017/27/L/ST9/03221. 

\bibliographystyle{aasjournal}
\bibliography{pap}

\clearpage

\begin{longrotatetable}
\begin{deluxetable}{lrrrrrrrl}
\tabletypesize{\footnotesize}
\tablecaption{OGLE and \textit{Gaia}~EDR3 proper motions of analyzed microlensing events (first 20 events).\label{tab:pm}}
\tablehead{
\colhead{Event} & \colhead{R.A.} & \colhead{Decl.} & \colhead{$\mu_{\alpha}^{\mathrm{OGLE}}$} & \colhead{$\mu_{\delta}^{\mathrm{OGLE}}$} & \colhead{$\mu_{\alpha}^{Gaia}$} & \colhead{$\mu_{\delta}^{Gaia}$} & \colhead{RUWE} & \colhead{Other names} \\ 
\colhead{} & \colhead{(deg)} & \colhead{(deg)} & \colhead{(mas\,yr$^{-1}$)} & \colhead{(mas\,yr$^{-1}$)} & \colhead{(mas\,yr$^{-1}$)} & \colhead{(mas\,yr$^{-1}$)} & \colhead{} & \colhead{OGLE-} 
} 
\startdata
BLG100.6.38608 & $267.44682$ & $-29.94507$ & $-5.61 \pm 1.10$ & $-7.71 \pm 3.20$ & \tabdots & \tabdots & \tabdots & 2004-BLG-131 \\
BLG101.1.189629 & $268.62158$ & $-30.05339$ & $-3.61 \pm 0.34$ & $-3.81 \pm 0.16$ & \tabdots & \tabdots & \tabdots & \tabdots \\
BLG102.7.44461 & $268.99768$ & $-29.64132$ & $-0.05 \pm 0.08$ & $-2.20 \pm 0.08$ & $-0.084 \pm 0.090$ & $-2.244 \pm 0.060$ & 1.01 & 2005-BLG-474,PAR-07 \\
BLG103.1.8650 & $269.14740$ & $-30.38714$ & \tabdots & \tabdots & \tabdots & \tabdots & \tabdots & 2005-BLG-143 \\
BLG103.2.137076 & $269.21318$ & $-30.14468$ & \tabdots & \tabdots & \tabdots & \tabdots & \tabdots & \tabdots \\
BLG103.4.61174 & $269.10570$ & $-29.88569$ & $-6.02 \pm 2.08$ & $-6.31 \pm 1.36$ & \tabdots & \tabdots & \tabdots & 2009-BLG-008 \\
BLG103.7.181467 & $269.07318$ & $-30.17691$ & $-1.23 \pm 0.36$ & $-3.28 \pm 0.32$ & \tabdots & \tabdots & \tabdots & \tabdots \\
BLG104.7.157693 & $269.62318$ & $-29.51509$ & $-5.90 \pm 1.00$ & $-4.06 \pm 0.30$ & $-4.316 \pm 0.658$ & $-3.590 \pm 0.430$ & 1.04 & 2006-BLG-366,PAR-54 \\
BLG105.1.139552 & $270.46734$ & $-29.66144$ & $-1.65 \pm 0.48$ & $-8.92 \pm 0.16$ & $-1.910 \pm 0.095$ & $-8.427 \pm 0.067$ & 1.33 & 2008-BLG-310 \\
BLG105.7.119541 & $270.39652$ & $-29.55708$ & \tabdots & \tabdots & \tabdots & \tabdots & \tabdots & 2006-BLG-377 \\
BLG121.3.60630 & $266.66075$ & $-34.75806$ & $-1.01 \pm 1.20$ & $-6.10 \pm 1.26$ & \tabdots & \tabdots & \tabdots & \tabdots \\
BLG121.8.163924 & $266.29775$ & $-34.98458$ & $-6.64 \pm 5.00$ & $1.03 \pm 5.80$ & \tabdots & \tabdots & \tabdots & 2005-BLG-363 \\
BLG122.5.142833 & $267.20035$ & $-34.54644$ & $-8.47 \pm 2.48$ & $-7.00 \pm 1.86$ & \tabdots & \tabdots & \tabdots & \tabdots \\
BLG122.5.173028 & $267.08529$ & $-34.51059$ & $-4.68 \pm 1.04$ & $-4.93 \pm 1.16$ & $-0.793 \pm 0.308$ & $-3.980 \pm 0.195$ & 1.30 & \tabdots \\
BLG122.7.92161 & $267.19680$ & $-34.87890$ & $1.46 \pm 2.48$ & $-4.54 \pm 2.76$ & \tabdots & \tabdots & \tabdots & \tabdots \\
BLG129.7.173817 & $265.85362$ & $-34.24060$ & $-3.65 \pm 0.84$ & $-4.80 \pm 0.86$ & $-2.499 \pm 1.379$ & $-4.505 \pm 1.098$ & 1.10 & 2005-BLG-165 \\
BLG130.5.98747 & $266.28940$ & $-33.94674$ & \tabdots & \tabdots & \tabdots & \tabdots & \tabdots & 2007-BLG-275 \\
BLG131.1.141016 & $267.33144$ & $-34.38358$ & $-2.22 \pm 0.74$ & $-5.73 \pm 0.30$ & $-0.045 \pm 0.158$ & $-4.874 \pm 0.099$ & 1.05 & \tabdots \\
BLG134.5.193547 & $269.24911$ & $-33.92100$ & $-1.48 \pm 0.46$ & $-5.83 \pm 0.20$ & $-1.192 \pm 0.134$ & $-5.370 \pm 0.095$ & 1.36 & 2005-BLG-061,PAR-42 \\
BLG138.1.192949 & $266.64771$ & $-33.77216$ & $-0.25 \pm 0.20$ & $-3.95 \pm 0.08$ & $-0.618 \pm 0.085$ & $-4.070 \pm 0.053$ & 1.74 & 2004-BLG-361,PAR-05 \\
\tabdots & \tabdots & \tabdots & \tabdots & \tabdots & \tabdots & \tabdots & \tabdots & \tabdots \\
\enddata
\tablecomments{This table is available in its entirety in machine-readable form.}
\end{deluxetable}
\end{longrotatetable}

\clearpage

\begin{deluxetable}{lrrr}
\tablecaption{The most likely mass and distance to the lens as well as the probability that the lens is dark (first 20 events).\label{tab:all}}
\tabletypesize{\small}
\tablehead{
\colhead{Event} & \colhead{$M$} & \colhead{$D_l$} & \colhead{$p$}\\ 
\colhead{} & \colhead{($M_{\odot}$)} & \colhead{(kpc)} & \colhead{}
} 
\startdata
BLG100.6.38608  & $3.05_{-1.92}^{+6.97}$ & $4.65_{-1.85}^{+1.82}$ & 0.983 \\
BLG101.1.189629 & $0.65_{-0.48}^{+0.42}$ & $1.50_{-0.51}^{+2.14}$ & 0.622 \\
BLG102.7.44461  & $0.89_{-0.43}^{+0.96}$ & $4.56_{-1.47}^{+1.23}$ & 0.467 \\
BLG103.1.8650   & $2.30_{-1.57}^{+6.03}$ & $5.68_{-2.22}^{+1.14}$ & 1.000 \\
BLG103.2.137076 & $1.15_{-0.84}^{+5.68}$ & $5.71_{-3.10}^{+1.06}$ & 0.984 \\
BLG103.4.61174  & $0.28_{-0.14}^{+0.20}$ & $0.69_{-0.27}^{+0.48}$ & 0.838 \\
BLG103.7.181467 & $0.76_{-0.34}^{+0.85}$ & $5.03_{-1.28}^{+1.15}$ & 0.912 \\
BLG104.7.157693 & $0.44_{-0.21}^{+0.26}$ & $1.32_{-0.49}^{+0.94}$ & 0.998 \\
BLG105.1.139552 & $2.43_{-1.28}^{+4.47}$ & $3.91_{-1.29}^{+2.23}$ & 0.960 \\
BLG105.7.119541 & $3.88_{-2.78}^{+9.63}$ & $6.15_{-1.00}^{+0.78}$ & 0.999 \\
BLG121.3.60630  & $0.98_{-0.58}^{+5.95}$ & $3.51_{-2.32}^{+3.11}$ & 0.860 \\
BLG121.8.163924 & $6.75_{-5.34}^{+26.05}$ & $4.74_{-2.19}^{+1.78}$ & 0.991 \\
BLG122.5.142833 & $0.37_{-0.16}^{+0.23}$ & $1.46_{-0.48}^{+0.60}$ & 0.985 \\
BLG122.5.173028 & $0.40_{-0.14}^{+0.21}$ & $1.84_{-0.53}^{+0.66}$ & 0.813 \\
BLG122.7.92161  & $1.23_{-0.82}^{+6.18}$ & $3.34_{-1.74}^{+2.58}$ & 0.943 \\
BLG129.7.173817 & $3.19_{-1.97}^{+5.33}$ & $5.65_{-1.71}^{+1.24}$ & 0.993 \\
BLG130.5.98747  & $2.25_{-1.70}^{+7.47}$ & $5.57_{-2.50}^{+1.30}$ & 0.989 \\
BLG131.1.141016 & $2.51_{-0.76}^{+0.98}$ & $3.58_{-0.62}^{+0.64}$ & 0.992 \\
BLG134.5.193547 & $0.35_{-0.16}^{+0.26}$ & $1.67_{-0.63}^{+0.79}$ & 0.383 \\
BLG138.1.192949 & $2.06_{-0.64}^{+0.87}$ & $4.04_{-0.71}^{+0.73}$ & 0.805 \\
\tabdots & \tabdots & \tabdots & \tabdots \\
\enddata
\tablecomments{This table is available in its entirety in machine-readable form. We provide the median and 68\% symmetric credible intervals on lens mass and distance.}
\end{deluxetable}

\end{document}